\documentclass[12pt]{article}
\usepackage{latexsym}
\usepackage[dvips]{graphicx}
\usepackage[english]{babel}
\usepackage{epsfig}
\newcommand \jp {$J/\psi \,$}
\newcommand \psip {$\psi' \,$}

\begin{document}
\title{Cross Section Oscillations in the Coherent 
Charmonium Photoproduction off Nuclei at Moderate Energies} 
\author{
L.~Frankfurt, L.~Gerland\\
\it School of Physics and Astronomy, Raymond and Beverly Sackler\\
\it Faculty of Exact Science, Tel Aviv University, Ramat Aviv 69978,\\
\it Tel Aviv, Israel\\
M.~Strikman\\
\it Pennsylvania State University, University Park, Pennsylvania 16802\\
M.~Zhalov\\
\it Petersburg Nuclear Physics Institute, Gatchina 188350, Russia}
\date{}
\maketitle
\centerline {\bf ABSTRACT}
We calculate the coherent charmonium photoproduction at intermediate
energies accounting for the physics of the charmonium bound states and the
dependence of the cross section on the region occupied by color using a
correspondingly adjusted generalized vector dominance model (GVDM).  In the
photon energy domain where the coherence lengths are comparable to the
average internucleon distances in nuclei and the nuclear radii we found that
significant oscillations of the total and forward photoproduction cross
sections governed by the longitudinal nuclear form factor are strongly
modified by the charmonium rescatterings accounting for the nondiagonal
transitions related to the color screening phenomenon.  We discuss how these
oscillations can influence the determination of the genuine
charmonium-nucleon cross sections in the forthcoming SLAC E160 experiment on
low energy \jp and \psip photoproduction off nuclei.
                                                                            
\section{Introduction}

One of striking QCD predictions is the dependence of the hadron interaction
strength on the volume occupied by color. This phenomenon has been suggested
initially within the constituent quark model of hadrons and the two gluon
exchange model for a hadron-hadron interaction~\cite{Low:1975sv, Gunion:iy}
and proved later for the interaction of spatially small quark-gluon wave 
packages as another form of QCD factorization
theorem~\cite{Blattel,rad}. The interaction of the charmonia with nuclei is
a natural field for exploring this idea. The probability of spatially small
c$\bar c$-configurations within these hadrons is large as follows from
the observed cross section of charmonium
photoproduction~\cite{FS91,FKS97,kophuf}. Within the charmonium models(see
e.g.~\cite{Buchmuller:zf}) the radius of the \psip is twice as large as the
radius of the \jp, hence, one can expect that their interaction with a
nucleus should be strongly different.  Strong activity in the search of the
quark-gluon plasma resulted in an enhanced interest to the accurate
treatment of charmonium-nucleus interactions. Really, the suppression of
the charmonia yield in the central ultrarelativistic heavy ion collisions
has been suggested
as one of the most promising signals of the Quark Gluon Plasma (QGP) 
(for a review see e.g.~\cite{Kharzeev:1995kz}). However, competing
suppression of the
charmonia yield is due to the ordinary final state charmonia interaction
with hadron matter - the interaction with target and projectile nucleons and
the effect of co-movers\cite{bm88}, hence, well determined \jp N an \psip
N cross sections are urgently needed. 
Obviously these quantities can't be found by direct
measurements. Also they cannot be extracted from the studies of the
photo/electro production off a nucleon  since 
average $c\bar c$  separation in this process is significantly smaller than 
the average size of oniums. Hence the only way is to extract them from the 
measurement of charmonia 
nuclear photo- and hadroproduction processes where \jp and \psip interact
with the nucleons during their passage through the nucleus from the
production point.  At present the uncertainties in the extracted cross
sections are too large - the values are ranging from 1 mb up to 8 mb for \jp
N and, correspondingly, from 0.8 mb to 20 mb for \psip N. This is due to
both the purely experimental problems and the
theoretical issues(for review and extensive
list of references see \cite{vmvogt}).

In the discussion of vector meson photoproduction one needs to distinguish
several different energy ranges which correspondingly require different
approximations. They are related to the energy dependence of the coherence
length $l_c\approx {{m_V^2}\over {2\omega}}$. At low energies the meson is 
formed at the distances smaller than
the typical interaction length of a meson in the nucleus. In this case one
should take into account the significant probability of multiple
rescattering  with different mesons in the intermediate state  
(i.e.\ $\gamma +A\to V^{\prime}+A\to V+A$) 
due to the presence of the
nondiagonal interactions of vector mesons with nucleons
($VN\to V'N$). The opposite limit is that of very high energies in which a
photon converts to a system of $q\bar q$ and higher 
Fock-components before the
target and one has to account for the interactions of this system with the
media and subsequent transition of the system to a vector meson. In the case
of the production of light mesons the soft physics dominates in both cases
and hence the use of the Gribov-Glauber model should be applicable in a wide
range of energies (for most of the cases the inelastic shadowing remains a
small correction and the Gribov approximation effectively reduces to the
eikonal model) As a result one can use the Vector Dominance Model(VDM)
 with nondiagonal transitions
 - the Generalized Vector Dominance Model (GDVM) to describe the $\rho, 
\rho'$ production in a wide range of energies, see e.g.~\cite{Shaw,FSZrho}.  

The situation is much more involved in the case of (charm)onium production.
At high energies both the coherence length and the formation length 
$l_f \approx 2\omega {[m_{\psi'}^2-m_{J/\psi}^2]}^{-1}$ (distance
on which the squeezed $q\bar q$ pair transforms into the ordinary meson)
are large and the color transparency
phenomenon reveals itself explaining the fast increase of the cross section
with energy observed at HERA (for a review see ref.~\cite{hera}). Then
the value of the cross section extracted from the charmonium photoproduction
characterizes the interaction of the squeezed $c\bar c$ pair with a 
nucleon rather than the charmonium-nucleon interaction.
In the high energy limit the very  small interquark distances in the 
wave function of the photon dominate and one has to treat
the interaction of small dipoles with nuclei. In this case the eikonal
approximation gives a qualitatively wrong answer since it does not take into
account the leading twist effect of the gluon shadowing(see the detailed
analysis in~\cite{FGMS2002}) while the leading twist analysis predicts large
shadowing effects~\cite{FSZpsi}. On the other hand at the intermediate
energies when onium states are formed inside the
nucleus the nonperturbative effects at a transverse distance scale
comparable to the charmonium size becomes important.  
Obviously, the hadronic basis description would be more relevant 
in this case but the VDM which is
grounded on such a basis failed to account properly for
the basic QCD dynamics of interaction.  In particular, SLAC
data~\cite{slac2} show that $0.15 \cdot \sigma_{\gamma+N\rightarrow \psi
N}\approx \sigma_{\gamma+N\rightarrow \psi'N}$ which within VDM corresponds
to:  $\sigma_{\psi' N} /\sigma_{J/\psi N}\approx 0.7$. This conclusion is in
evident contradiction with the QCD expectation where the hadron
interaction depends on the volume occupied by color and the cross sections
should be scaled approximately as the transverse area occupied by color:
$\sigma_{\psi' N}/\sigma_{J/\psi N} \propto r_{\psi'}^2 /r_{J/\psi}^2$. So
within this naive QCD picture the cross section of the \psip N interaction
could be larger by a factor of the order of four. A QCD explanation of such
failure of VDM is based on the
observation\cite{FS91,kophuf} that in photoproduction of both 
the \jp and \psip mesons the small relative distances
$\sim 1/m_c$ dominate in the $c\bar c$ component of the photon wave function.
As a result the relative suppression of the \psip production
as compared to the \jp production is primarely related to a larger
nonconservation of mass in the $\gamma\rightarrow \psi'$ transition and is
of the order of $m_{\psi}^2/m_{\psi'}^2$. 
Overall the $J/\psi,\,\psi'$  photoproduction data strongly 
indicate that the dependence of the interaction on the size of the 
region occupied by color takes place already at moderate energies.
The dominance of small
$c\bar c$ configurations in photoproduction processes is relevant for
the significant probability of nondiagonal 
$J/\psi \leftrightarrow  \psi ^\prime$ diffractive transitions and
the proper account for QCD dynamics within a hadronic basis 
can be provided by using the
Generalized Vector Dominance Model (GVDM) adjusted to account for 
the color screening phenomenon~\cite{FS91,kophuf,knnz} 
in the regime of small coherence lengths where leading twist
shadowing is not important.
 
In this paper we use the GVDM to consider the coherent photoproduction
of hidden charm mesons off nuclei at moderate photon energies $20\mbox{
GeV} \leq \omega \leq 60\mbox{ GeV}$ where the coherence length for the
$\gamma V$ transition $l_{c}$ is
still close enough to the internucleon distance in nuclei while the
formation length $l_{f}$ is comparable to the radii of heavy nuclei (the
problem of interpolating between this regime and the regime of the leading
twist gluon shadowing dominance is beyond the scope of this paper). In this
energy range it is reasonable to expect that, at least, with a heavy nucleus
target there is a noticeable probability for rescattering of charmonia, hence,
one would be able to reveal the fluctuation of the charmonium-nucleon
interaction strength as due to the diagonal $\psi N\rightarrow \psi N$ and
nondiagonal $\psi N\Leftrightarrow \psi' N$ rescatterings. We would like to
emphasize that at moderate energies using the Generalized Vector Dominance
Model (GVDM) whose parameters are chosen to guarantee the validity of color
screening phenomena allows to account for the space-time evolution of
spatially small $c\bar c$ pair together with the Glauber model approximation
generalized to account for the physics related to the coherence length. It is
important that the inelastic shadowing corrections related to the
production of higher mass states~\cite{Gribovinel} are still
insignificant.

We built an approach, the Generalized Glauber Model (GGM), based on
combining of the multistep production Glauber model with the GVDM. Within
this approach we perform calculations aimed to investigate how the color
fluctuations reveal themselves in the interactions of charmonium 
states with nuclear medium in coherent photoproduction off nuclei. 
It should be noted that the charmonium photo(electro)production  off
nuclei in the wide range of the photon energies  is a subject of active
investigation(see for example \cite{jct}). However, as far as we find 
this paper for the first time raises and investigates the question of the
significant cross section oscillations  in the total and forward cross
sections of the \jp and \psip photoproduction 
off nuclei at low and moderate energies of photons. 
Our analysis indicates that this new observation should be very useful
and important for the experimental measurement (E160) of the charmonia
photoproduction planned at SLAC~\cite{SLAC} 
to obtain a reliable experimental estimate for  
the genuine cross sections of $J/\psi N$ and $\psi' N$.

\section{Description of the model} 

Our interest in this paper is in moderate energy range phenomena where the
$c\bar c$ configuration is spatially small in the production point.  The
large mass of $c$ quark is important to ensure the applicability of PQCD for
the description of photoproduction the processes and the interaction with
target of this initial $c\bar c$ configuration. However, the spatially small
configuration transforms into a hadron state before it can collide with a
second nucleon. 
So a more accurate treatment  of  the photoproduction cross section 
\begin{eqnarray}
\sigma_{\gamma A\to V A}(\omega)  =\int \limits_{-\infty}^{t_{min}} {\rm d}t
\frac {\pi} {k_{V}^2}{\left |F_{\gamma A\to VA}(t)\right |}^2=\frac {\pi} {k_{V}^2}
\int \limits_{0}^{\infty} {\rm d}t_{\bot }
{\left|\frac {ik_{V}} {2\pi}  \int \,{\rm d}\,{\vec b}\,  e^{i{\vec q}_{\bot 
}\cdot {\vec b}}
\Gamma _{V} ({\vec b}) 
\right|}^2\quad.
\label{crosec}
\end{eqnarray}
can be achieved within a hadronic basis. Here $V=J/\psi,\Psi',\cdots$,
${\vec q}\,^2_{\bot }=t_{\bot }={t_{min}-t}$, 
$-t_{min}=\frac {M_{V }^4} {4\omega^2}$ 
is the longitudinal momentum transfer in the $\gamma \to V $ transition, 
and $\Gamma _{V}({\vec b})$ is the diffractive nuclear profile function
\begin{eqnarray}
\Gamma _{V}({\vec b})=\lim _{z\rightarrow \infty}\Phi _{V}({\vec b},z)\quad.
\label{gamma}
\end{eqnarray}
The  eikonal function $\Phi _{V}({\vec b},z)$ takes into account 
the phase difference between the incident  photon and the 
intermediate resp.\ final states due to the longitudinal momentum 
transfer.
It is calculated in the Generalized Glauber model where the 
modified Glauber approach formulae~\cite{bochmann} were combined with the
GVDM~\cite{GVDM,GVDM1}. 

A reasonable  starting approximation to evaluate the amplitude of the
charmonium-nucleon  interaction is to restrict ourselves to the basis of
\jp and \psip states for the photon wave function. Then
\begin{eqnarray}
f_{\gamma N \to J/\psi N}=\frac {e} {f_{J/\psi }} f_{J/\psi N\to J/\psi N}+
\frac {e} {f_{\psi'}}f_{\psi' N\to J/\psi N}\quad, 
\nonumber
\\
f_{\gamma N\to \psi' N}=
\frac {e} {f_{\psi'}} f_{\psi' N\to \psi' N}+
\frac {e} {f_{J/\psi }}f_{J/\psi N\to \psi' N}\quad .
\label{gvdm}
\end{eqnarray}

Here the coupling constants $f_{V}$ are determined from the widths 
of the vector meson decays $V \rightarrow e\bar e$
\begin{equation}
\left({e^2\over 4 \pi f_{V}}\right)^2={3\over 4\pi} 
{\Gamma(V\rightarrow e\bar e)\over m_{V}}\quad.
\end{equation}
with~\cite{pdg}
\begin{eqnarray}
\Gamma(\psi\rightarrow e\bar e)=5.26\pm 0.37\mbox{ keV}\mbox{ and }
\Gamma(\psi'\rightarrow e\bar e)=2.12\pm 0.18\mbox{ keV}\quad.
\end{eqnarray}
This yields
\begin{eqnarray}
{f_{\psi}^2\over 4\pi}=10.5\pm0.7\mbox{ and }
{f_{\psi'}^2\over 4\pi}=30.9\pm2.6\quad.
\end{eqnarray}

In the optical limit ($A\gg 1$) the generalized Glauber-based optical
potentials in the short-range approximation are given by the expression
\begin{eqnarray}
U_{iA\to jA}({\vec b},z)=-4\pi f_{iN\to jN}\varrho ({\vec b},z)\quad.
\end{eqnarray} 
where  $f_{iN\to jN}$ are elementary forward amplitudes and the nuclear density 
$\varrho (\vec b,z) $ is normalized by the condition
$\int d^{2}\vec b dz\,\varrho (\vec b,z)=A$. Thus in the optical limit, with
accuracy ${\cal O} (\sqrt \alpha_{em})$ the eikonal functions 
$\Phi_{J/\psi,\,\psi'} ({\vec b},z)$ are determined as the solutions of
the coupled two-channel equations
\begin{eqnarray}
2ik_{J/\psi } \frac {d} {dz} \Phi _{J/\psi }({\vec b},z)=& &
U_{\gamma A\to J/\psi A}({\vec b},z)
e^{iz\cdot q_{\| }^{\gamma J/\psi }}+
U_{J/\psi A\to J/\psi A}({\vec b},z)\Phi _{J/\psi }({\vec b},z)+
\nonumber
\\ 
&+&
U_{J/\psi A\to \psi' A}({\vec b},z)
e^{iz\cdot q_{\|}^{J/\psi \psi'}}\Phi _{\psi'}({\vec b},z)\quad ,
\label{eqjpsi}
\end{eqnarray}
\begin{eqnarray}
2ik_{\psi'} \frac {d} {dz} \Phi _{\psi'}({\vec b},z)=& &
U_{\gamma A\to \psi' A}({\vec b},z)
e^{iz\cdot q_{\| }^{\gamma \psi'}}+
U_{\psi' A\to \psi' A}({\vec b},z)\Phi _{\psi'}({\vec b},z)+
\nonumber
\\ 
&+&
U_{\psi' A\to J/\psi A}({\vec b},z)
e^{iz\cdot q_{\| }^{\psi' J/\psi }}\Phi _{J/\psi }({\vec b},z)\quad,
\label{eqpsip} 
\end{eqnarray}
with the initial condition $\Phi _{J/\psi ,\psi'} (\vec b ,-\infty)=0$. 
The exponential factors ${\rm exp} [iq_{\| }^{i\rightarrow j}z]$
account for the dependence of the  amplitudes on $t_{min}$. They 
are responsible for the coherent
length effect, $i,j=\gamma ,J/\psi \mbox{ resp. }\psi'$, $q_{\|}^
{i\rightarrow j}=\frac {M_j^2-M_i^2} {2\omega }$.
We calculated $\varrho (\vec b,z)$ in the Hartree-Fock-Skyrme (HFS)  model
which provided a very good (with an accuracy $\approx 2\%$) description of
the global nuclear properties of spherical nuclei along the periodical table
from carbon to uranium~\cite{HFS} and the shell momentum distributions in
the high energy (p,2p)~\cite{p2p} and (e,e'p)~\cite{eep} reactions.
   
The distinctive feature of the coherent charmonia photoproduction off nuclei
at moderate energies is that the amplitude for nondiagonal transitions is
large. As an educated guess it can be estimated from the observation that in
the photoproduction processes $c\bar c$-pairs are produced within a
spatially small configuration. Within the framework of
charmonium models the overlapping integral between the wave functions of
photons and mesons with hidden charm is proportional to the charmonium wave
function at zero distances. Thus at not too large energies the interaction
with a nucleon of such small $c\bar c$ configuration should be strongly
suppressed by the small transverse area occupied by color. If one neglects
for the moment the relatively small photoproduction amplitude, one obtains
from eq.~(\ref{gvdm}) for the nondiagonal amplitude
\begin{eqnarray}
f_{\psi' N\to J/\psi N}
\approx -{f_{\psi'}\over f_{J/\psi}}\cdot f_{J/\psi N\to J/\psi N} 
\approx -1.7\cdot f_{J/\psi N\to J/\psi N}\quad. 
\end{eqnarray}
and an even larger value for $\psi'N$ interaction
\begin{eqnarray}
f_{\psi' N\to \psi' N}\approx {f^2_{\psi' }\over f^2_{J/\psi}}\cdot 
f_{J/\psi N\to J/\psi N}\approx 3 \cdot f_{J/\psi N\to J/\psi N }\quad.
\end{eqnarray}

Data on $\psi'$ absorption in nucleus-nucleus collisions suggest the value
of $\sigma_{\psi' N}\sim$ 20 mb~\cite{SA} which combined with the SLAC
data~\cite{slac1} corresponds to $\sigma_{\psi' N}/\sigma_{J/\psi N}\approx
5\div 6$ with large experimental and theoretical errors. Large values for
nondiagonal amplitudes are a characteristic QCD property of hidden charm and
beauty meson-nucleon interaction. Note that the negative sign of the
nondiagonal amplitude is dictated by the QCD factorization theorem. A
positive sign of the forward photoproduction $f_{\gamma N\rightarrow
J/\psi(\psi') N}$ amplitudes as well as the signs of the coupling constants
$f_{J/\psi}$ and $f_{\psi'}$ are determined by the signs of the charmonium
wave functions at r=0. In order to fix the elementary amplitudes in the GVDM
more accurately we have used the following logic. We parametrize the cross
section in the form used by the experimentalists of HERA to describe their
data. This form has no firm theoretical justification but it is convenient
for the fit 
\begin{equation} 
\sigma_{\gamma+N \rightarrow V+N} \propto F_{2g}^2 \left({s\over 
s_0}\right)^{2\lambda}\quad. 
\end{equation} 
Here $s=2\omega m_N +m_N^2$ is the center-of-mass energy, $F_{2g}$
is the two-gluon form factor of a nucleon and $\lambda$ is derived from
experimental data~\cite{slac2} 
\begin{eqnarray} 
\left.{{\rm d}\sigma_{\gamma N\rightarrow J/\psi N}\over {\rm d}t}
\right|_{t-t_{min}=0}&=&17.8\pm1.5\mbox{ nb}\mbox{GeV}^{-2}\quad \mbox{at 
}s_0=40.4\,
\mbox{GeV}^{2}\quad ,\cr \left.{{\rm d}\sigma_{\gamma N\rightarrow J/\psi N} 
\over {\rm d}t}
\right|_{t-t_{min}=0}&=&40\pm13\mbox{ nb}\mbox{GeV}^{-2}\quad
\mbox{at }s_0=188.9\,\mbox{GeV}^{2}\quad, 
\end{eqnarray} 
and 
\begin{equation}
\left.{{\rm d}\sigma_{\gamma N\rightarrow \psi' N}\over 
{\rm d}t}\right|_{t-t_{min}=0}=
0.15\cdot\left.{{\rm d}\sigma_{\gamma N\rightarrow J/\psi N}\over {\rm d}t}
\right|_{t-t_{min}=0}\quad, 
\end{equation} 
and practically doesn't depend (or depends only weakly) on the energy.  The
two gluon form factor as extracted from analysis of the $J/\psi$
photoproduction data in~\cite{Frankfurt:2002ka} can be used to evaluate the
nucleon form factor at $t=t_{min}$: 
\begin{equation} 
F_{2g}={\left(1-{t_{min}\over m^2_{2g}}\right)^{-2}}\quad. 
\end{equation} 
The quantity $m^2_{2g}$ is defined by 
\begin{equation} 
{1\over m^2_{2g}}={1\over{\rm GeV}^2}+{0.06\over{\rm  
GeV}^2}\ln\left({s\over s_{0}}\right)\quad.
\end{equation} 
As a result for the photoproduction of the $J/\psi$ we found $\lambda=0.2$
while the value for the $\psi'$ production is somewhat less $\lambda=0.15$
but with much larger uncertainties. Hence, for the preliminary estimates one
can use the same value of $\lambda$ for both processes.

The $\psi N$ cross section can be parametrized as the sum of soft and 
hard physics:
\begin{equation}
{\sigma_{J/\psi N}(s)\over \sigma_{J/\psi N}(s_{0})}
=c\left({s\over s_{0}}\right)^{0.08} +(1-c) \left({s\over
s_{0}}\right)^{\lambda}\quad.
\end{equation} 
 
The SLAC data~\cite{slac1} shows that at $s_0=38.5$~GeV${}^2$
\begin{equation}
\sigma_{J/\psi N}(s=s_{0})=3.5\pm0.8 \mbox{ mb}\quad.
\end{equation}

We evaluated $c\approx 1$ in our previous paper on the absorption of 
$\psi$ produced in AA collisions~\cite{ger} with

\begin{equation}
\sigma(hard)=2\pi\cdot \int_{0.1 {\rm fm}}^{0.2 {\rm fm}} |\phi(b,z)|^2\cdot 
\sigma(b) b\,{\rm d}b\, {\rm d}z\quad.
 \end{equation}
Here
$\phi(b,z)$ is the wave function of the $\psi$, $\sigma(b)$ the 
perturbative dipole 
cross section from~\cite{rad}. The limits of the integration are for the $b$ 
integration, the integration over the longitudinal direction $z$ is 
from $-\infty$ to $+\infty$.

When the upper limit of the integral is increased to 0.35 fm then $c=0.915$,
i.e. $\sigma(hard)=(1-c)\cdot \sigma_{J/\psi N}(s=s_{0})=0.3$ mb. This is an
uncertainty of the model since it is not clear up to which value of $b$ PQCD
is applicable. However, in the following $c=0.915$ is used. That is
\begin{equation}
\sigma_{J/\psi N}=3.2\mbox{ mb}\left({s\over s_{0}}\right)^{0.08} +
0.3\mbox{ mb} 
\left({s\over s_{0}}\right)^{0.2}\quad.
\label{psiN}
\end{equation}

The existence of a hard part of the $J/\psi$N cross section is consistent
with the GVDM, because the photoproduction amplitude has a stronger energy
dependence than the Pomeron exchange, i.e.\ soft scattering amplitudes. 
size meson is not surprising, since
pQCD amplitudes for colorless dipoles have an energy dependence similar to
the photoproduction amplitude due to the energy dependence of the gluon
density distribution. Therefore the energy dependence of the dipole nucleon
cross section should gradually increase with decreasing size of the dipole
(the meson), otherwise there would be a jump in the energy dependence from
soft to hard processes. The energy dependence of the total cross section of
the charmonium-nucleon interaction as described by eq.~(\ref{psiN}) differs 
from that in \cite{kophuf} who assumed a dominance of pQCD and therefore 
a significantly faster increase of this cross section with energy.

Now one can find the imaginary part of the amplitude from the optical 
theorem\\ 
$\Im f_{J/\psi N\rightarrow J/\psi N}=s\sigma_{J/\psi N}$ and the real part 
using the Gribov-Migdal relation 
\begin{equation}
{\Re f_{J/\psi N\rightarrow  J/\psi N}}
={s\pi\over 2}{{\partial \over \partial 
\ln{s}}{\Im f_{J/\psi N\rightarrow J/\psi N}\over s}}\quad.
\end{equation}
\begin{figure}
    \begin{center}
        \leavevmode
        \epsfxsize=1.\hsize
        \epsfbox{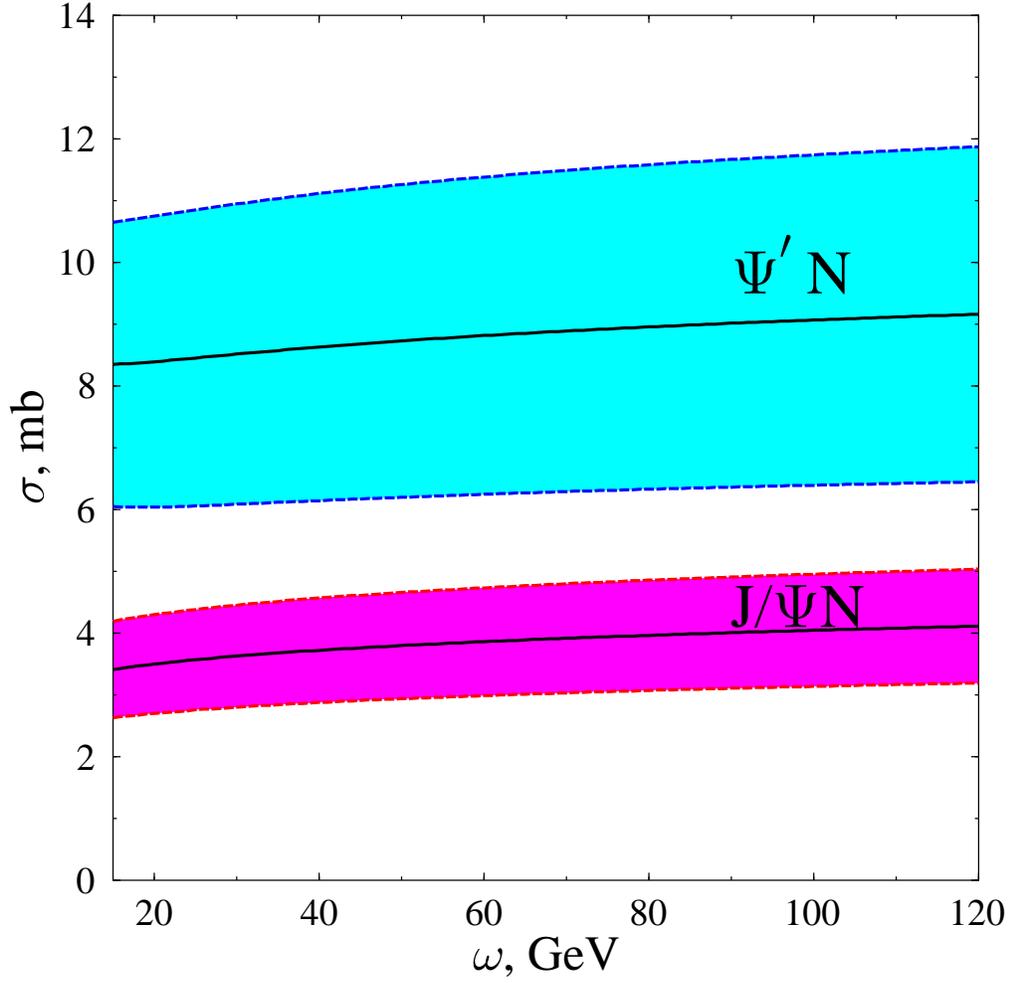}
    \end{center}
\caption{The energy dependence of the elementary charmonium-nucleon
cross sections found in the GVDM. The filled areas show the variation
of the cross sections due to the uncertainty of the experimental $J/\psi$N 
cross section.}
\label{elcs}
\end{figure}

Since the amplitudes of the photoproduction and the $J/\psi N$ diagonal
interaction are fixed we find all other forward amplitudes 
from the GVDM equations (\ref{gvdm}). As a result we have determined
all parameters of the model. In particular, we found the value
$\sigma_{\psi' N}\approx 8 $ mb. 
However, the experimental cross sections of the forward elementary 
photoproduction and, especially, the value of $\sigma_{J/\psi N}$, used
as input of the GVDM are known with large uncertainties. We checked
how a variation of $\sigma_{J/\psi N}$ within the experimental errors
will influence results of our calculations. The ranges of the
elementary $J/\psi N$ and $\psi' N$ cross sections which we have got 
within this procedure are shown in fig.~\ref{elcs}.

\section{Results and discussion}

First of all let us discuss one of the key approximations we have used in
our calculations. We build the GVDM restricting the basis by 
accounting for only two  $\psi$ meson states - the ground $1S$ state
($J/\psi$) and the lowest excited $2S$ state ($\psi ^{\prime}$).  
Since the coherent production of heavy charmed
states off nuclei at low energies is suppressed by the target form
factor it seems to be quite reasonable. However 
the presence of two nearly degenerate states 
$\psi'$ ($M_{\psi ^\prime}=3.686$ GeV)
 and  $\psi''$ ($M_{\psi ^{\prime \prime}}=3.77$ GeV) 
requires a separate treatment. 
The following discussion heavily uses the fact 
that the properties of the $\psi''$ are well described (actually has been
predicted) on the basis of the charmonium model by Eichten 
and collaborators~\cite{eichten,Richard:1979fc}. Within this approach they 
have found that the $\psi''$ is the ${}^3D_1$ 
state with a small admixture of the $S$ wave and correspondingly $\psi'$
has a small admixture of the $D$ wave. Namely
\begin{eqnarray}
\left|\psi'\right>&=&\cos \theta  \left|2S\right> + \sin\theta 
\left|1D\right>\quad,\cr
\left|\psi''\right>&=&\cos \theta  \left|1D\right> - \sin\theta 
\left|2S\right>\quad.
\end{eqnarray}
Since only the S-wave contributes to the decay 
of $\psi$ states into $e^+e^-$ 
(at least in the nonrelativistic charmonium models)
the value of $\theta $ can be determined from the 
data on the $e^+e^-$ decay widths  $\Gamma (\psi'\rightarrow e^{+}e^{-})=2.14$ KeV
 and the $\Gamma (\psi''\rightarrow e^{+}e^{-})=0.26$ KeV as
\begin{equation}
\tan^{2}\theta=
{\Gamma (\psi''\rightarrow e\bar e)\over \Gamma(\psi'\rightarrow e\bar 
e)}\approx 0.1\,\Rightarrow\,
\theta =19^o \pm 2^o\quad.  
\end{equation}
Due to the small difference of masses between the $\psi ^\prime$ and $\psi ^{\prime \prime}$
mesons the produced $c\bar c $-state
corresponding to the S-wave does not loose coherence while going through the
media at any conceivable energies\footnote{
Interesting enough is that the mixing model allows us to predict the ratio 
of the $\psi'$ and $\psi''$ production cross section
in hard processes. It is
$$
{\sigma(\psi'')\over\sigma(\psi')}=\tan^2 (\theta) \approx 0.1\quad.
$$
This ratio should be a universal number for a hard processes, almost
independent on the process, hence, can be used for the cross
check whether charm production in the pA and in AA collisions is dominated
by hard processes.}. Also the soft  interactions cannot
transform the $S$-state to $D$-state with any significant probability. 
In the soft QCD processes data show that the cross sections
of exclusive nondiagonal transitions are negligible for the forward angle 
scattering. The same conclusion is valid in the PQCD model for the
charm dipole-nucleon interactions.
Hence instead  of writing a two-state matrix with  $J/\psi$ and $\psi'$
states it would be more appropriate to use the $1S$ - $2S$ 
basis. The only changes we would encounter would be 
the necessity of a change of $f_V$ since the  amplitude of the $2S$ state
production is larger by a factor of $1/\cos\theta$ than the 
amplitude of the $\psi'$  production. This yields
a $\sim 10\%$ effect which is within the uncertainties of the model.
In principle, one should account for the other charmonium states 
and such extension of the basis
could lead to some renormalization of the elementary amplitude
 but should not influence strongly coherent processes off nuclei
because contributions of the higher mass states are suppressed by the nuclear 
form factor. 
\begin{figure}
    \begin{center}
        \leavevmode
        \epsfxsize=1.\hsize
        \epsfbox{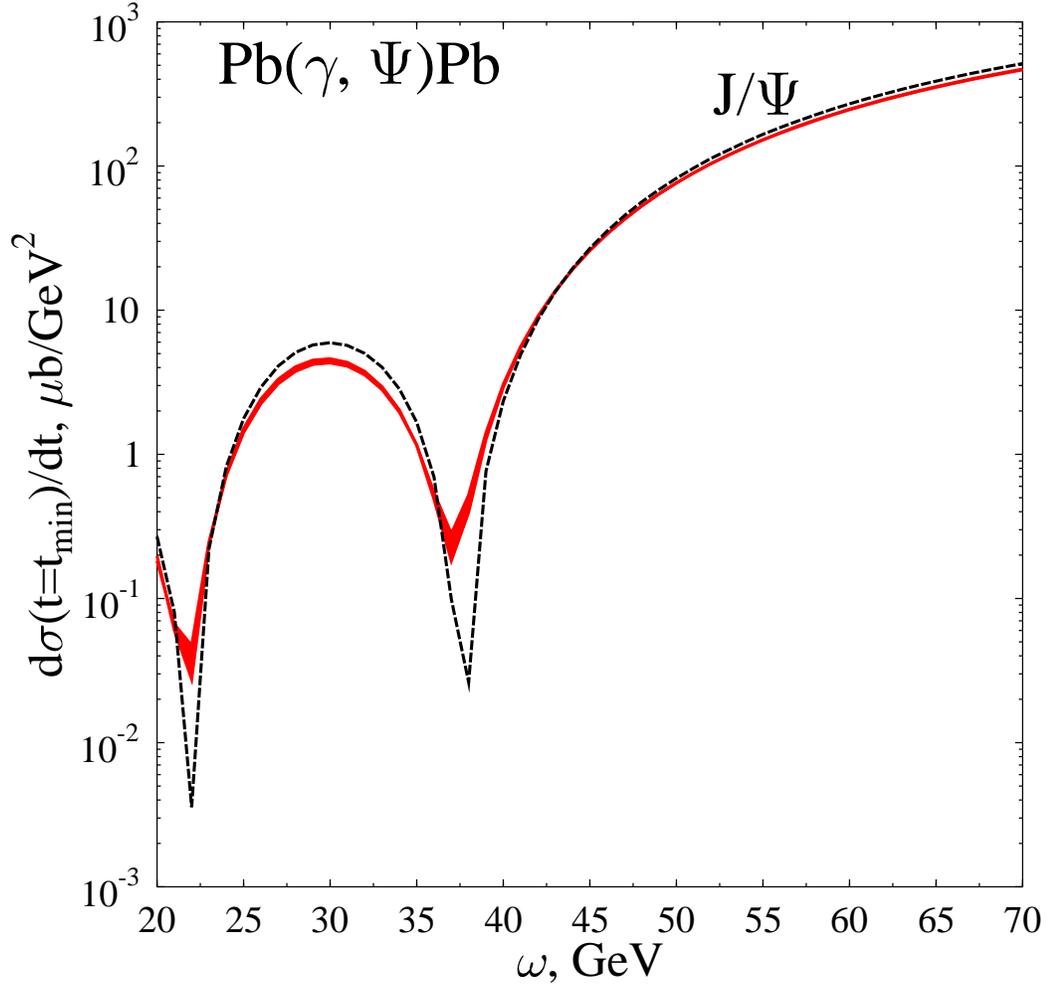}
    \end{center}
\caption{The energy dependence of the forward coherent $Pb(\gamma ,J/\psi)Pb$
cross section  calculated in the Generalized Glauber Model compared to the 
cross section in the Impulse Approximation (dashed line). The filled area
depicts the uncertainty due to the uncertainty of the elementary cross
sections shown in fig.~\ref{elcs}.}
\label{forjpsi}
\end{figure}

\begin{figure}
    \begin{center}
        \leavevmode
        \epsfxsize=1.\hsize
        \epsfbox{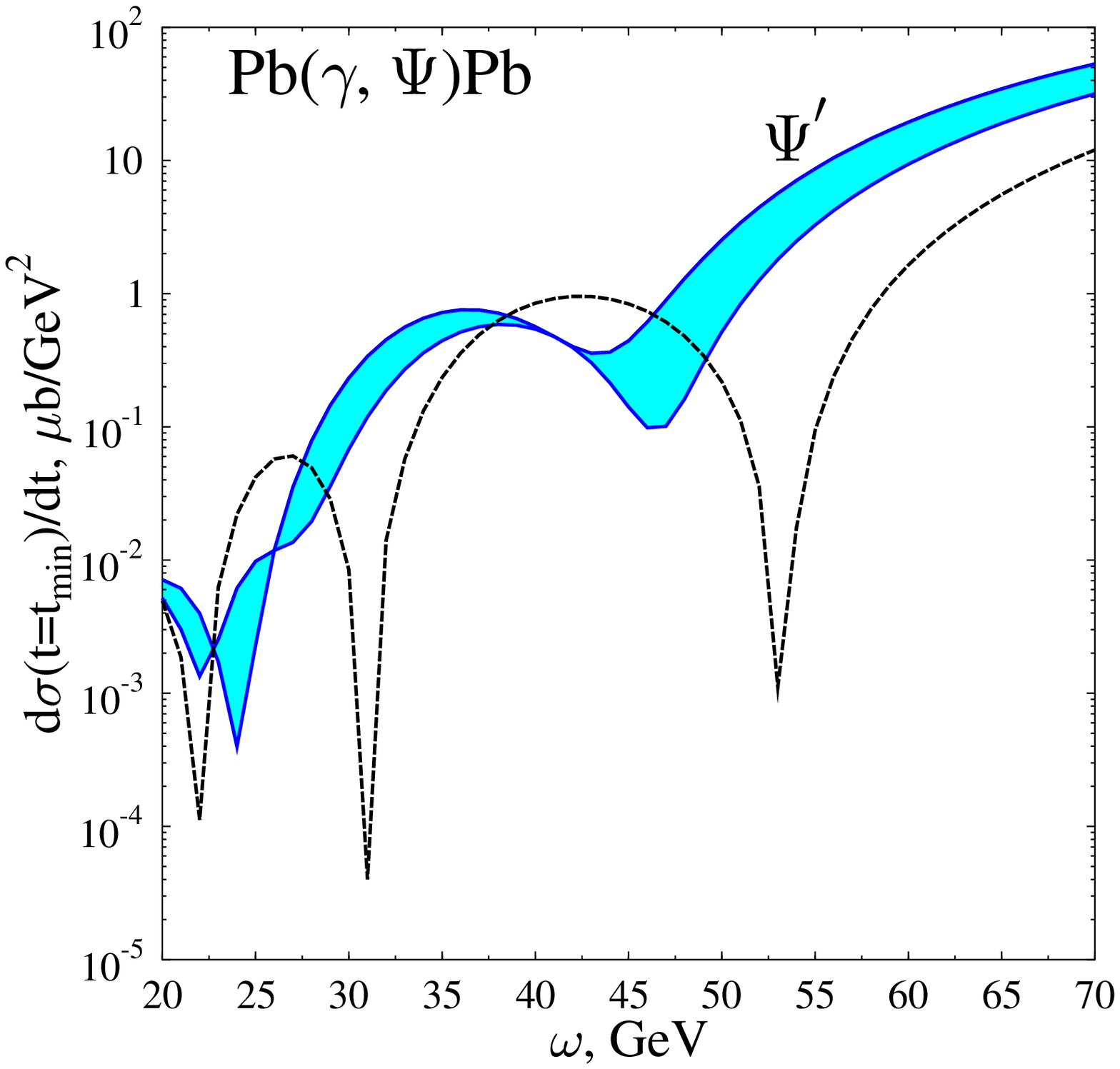}
    \end{center}
\caption{The energy dependence of the forward coherent 
$Pb(\gamma ,\psi^{\prime})Pb$  cross section calculated in the Generalized
Glauber Model compared to the cross section in the Impulse Approximation
(dashed line). The filled area depicts the uncertainty due to the 
uncertainty of the elementary cross sections shown in fig.~\ref{elcs}.
}
\label{forpsi}
\end{figure}

\begin{figure}
    \begin{center}
        \leavevmode
        \epsfxsize=1.\hsize
        \epsfbox{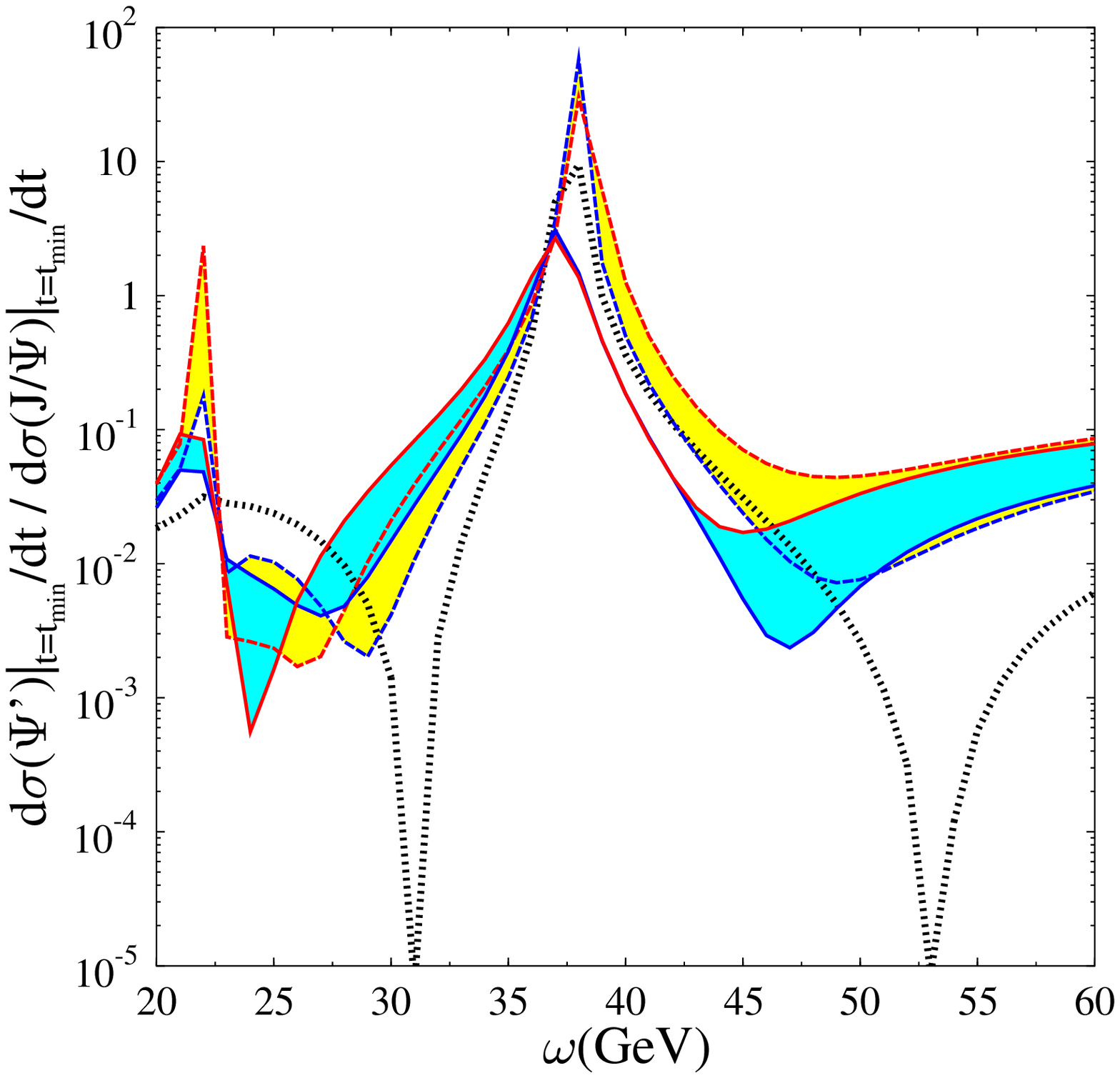}
    \end{center}
\caption{
The energy dependence of the ratio of the forward coherent charmonium 
photoproduction cross sections calculated in the Generalized Glauber 
Model(light-blue, solid lines)
compared to the ratio of cross sections in the Impulse Approximation 
(black dotted line) and in the  GGM with diagonal amplitudes
$f_{VN\to VN}=0$ (yellow, dashed lines). The filled areas
depict the uncertainty due to the uncertainty of the $J/\psi N$ 
elementary cross section shown in fig.~\ref{elcs}.
}
\label{ratfor}
\end{figure}

Within the model described in the previous section we calculated the 
forward cross
sections of the coherent photoproduction of $J/\psi$ and $\psi'$ off lead.  
The energy dependence of these cross sections is compared 
(Figs.\ref{forjpsi} and \ref{forpsi})  
to that obtained in the Impulse Approximation where all rescatterings 
of the produced vector mesons are neglected and the cross section is given
by the simple formula
\begin{eqnarray}
{d\sigma_{\gamma A\to VA}(\omega_{\gamma})\over {d\,t}}=
{d\sigma_{\gamma N\to VN}(t_{min})\over dt}\cdot
{\left |
{\int \limits_{0}^{\infty}e^{i{\vec q}_{\bot }\cdot \,{\vec b}}
{\rm d}\,{\vec b}\int \limits_{-\infty}^{\infty}{\rm d}\,z\,
e^{iz\cdot q_{\| }^{\gamma V }}\varrho  ({\vec b},z)}
\right |}^2\quad.
\label{iacs}
\end{eqnarray}

Remember that $\sqrt{-t_{min}}=q_{\| }^{\gamma V}=m_V^2/2\omega$,
hence, decrease of the photon energy corresponds to increasing of the
longitudinal momentum transfer in the coherent photoproduction off nuclei. 
Note that we neglected the transverse momentum transfer dependence of the
elementary amplitudes since the dominating effects are determined by the
dependence of the nuclear form factor on $t_{\bot }$(influence of the
t-dependence of the  elementary amplitudes will be considered in the
forthcoming papers).

\begin{figure}
    \begin{center}
        \leavevmode
        \epsfxsize=1.\hsize
        \epsfbox{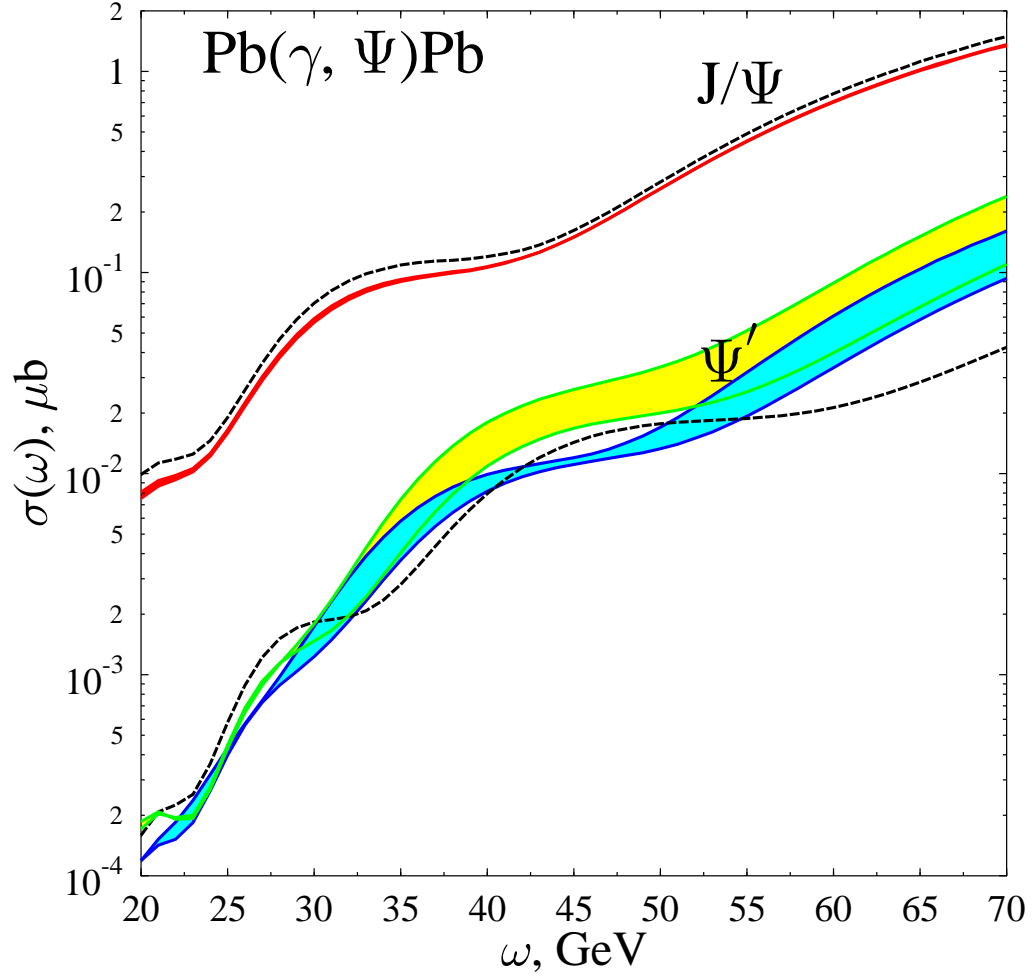}
    \end{center}
\caption{The energy dependence of the coherent charmonium photoproduction
cross sections  calculated in the Generalized Glauber Model
compared to the cross sections in the Impulse Approximation(dashed line)
 and in GGM with $f_{VN\to VN}=0$ (yellow). The filled areas
depict the uncertainty due to the uncertainty of the elementary cross
sections shown in fig.~\ref{elcs}.}
\label{cs}
\end{figure}

\begin{figure}
    \begin{center}
%        \leavevmode
        \epsfxsize=1.\hsize
        \epsfbox{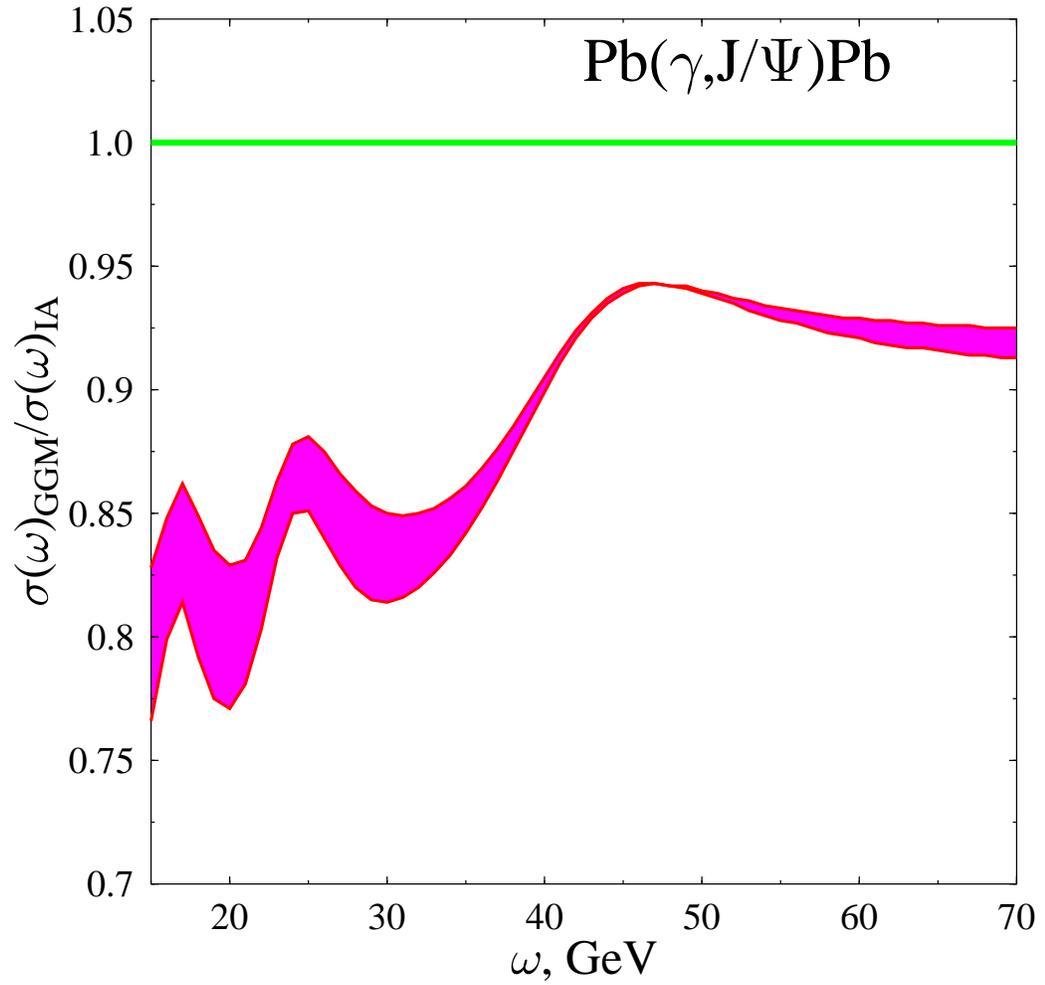}
    \end{center}
\caption{The energy dependence of the ratio of the coherent 
$Pb(\gamma ,J/\psi)Pb$ 
cross section calculated in the Generalized Glauber Model
to the cross section in the Impulse Approximation. 
The red filled area depicts uncertainty due to the variation of
the $J/\psi N$ experimental cross section.
}
\label{ratjpsi}
\end{figure}

\begin{figure}
    \begin{center}
%        \leavevmode
        \epsfxsize=1.\hsize
        \epsfbox{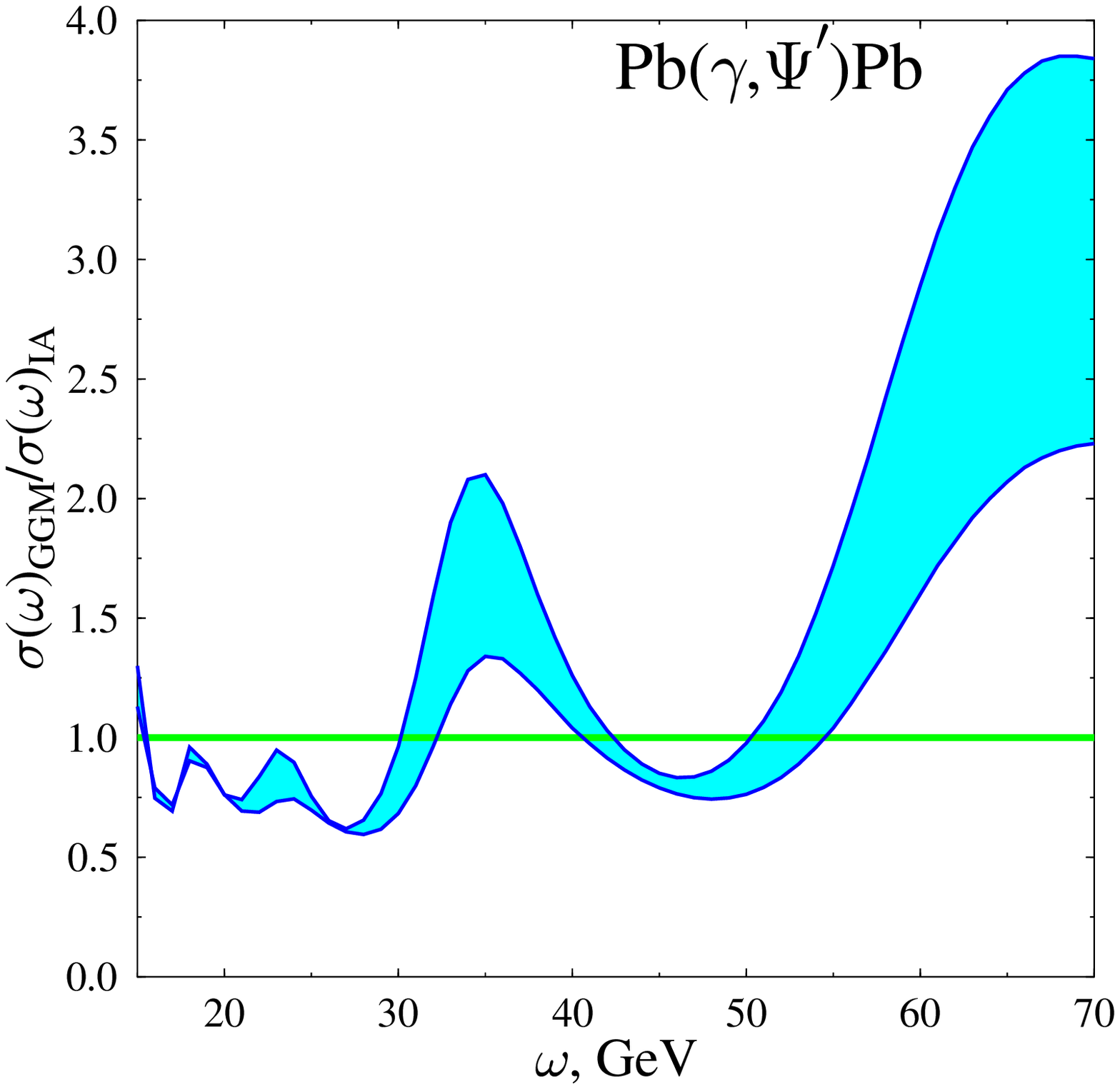}
    \end{center}
\caption{The energy dependence of the ratio of the coherent  
$Pb(\gamma ,\psi^{\prime})Pb$
cross section calculated in the Generalized Glauber Model
to the cross section in the Impulse Approximation. The filled light blue
area shows the uncertainty of the results due to the uncertainty in the
$J/\psi N$ cross section.
}
\label{ratpsi}
\end{figure}

The cross section of the \jp photoproduction in the GGM is close to
that calculated in the Impulse Approximation - the shapes of
curves are very similar and the values of the cross sections are slightly
reduced at energies below 40 GeV. A more striking difference is found
comparing the two approaches for the yield of \psip where both the shape and
the value of the cross section are essentially modified. 
The distinctive feature of the coherent charmonium
photoproduction cross sections is their oscillating behavior with the photon
energy. The major source for such a behavior is due to the oscillating
longitudinal nuclear form factor at the relatively large value of $t_{min}$ in 
the photoproduction vertex in the kinematical region we are 
interested in. One can easily check that the positions of
the first minimum satisfy the relation
$qR_A\approx 3.8$ well known from diffraction.  
We want emphasize that the use of a realistic nuclear
density in our calculations ensures a reasonable description of the nuclear
form factor. No such oscillations exist in the case of frequently used
Gaussian form for the nuclear form factors. From the comparison of 
calculations in
the GGM and in the IA it is seen that the rescattering don't change noticeably 
the cross section of the coherent \jp
photoproduction off nuclei except for the filling of the minima. 
All effects of the \jp N rescattering are obviously small compared to
the direct photoproduction and the longitudinal form factor is
weakly distorted by the final state \jp -nucleus interaction.
Hence, analyzing the coherent \jp
photoproduction off the spherical nuclei one can use 
the nuclear form factor determined from the study of the high energy elastic
electron-nucleus scattering.
The picture is qualitatively different for the coherent \psip
photoproduction off nuclei. Due to the higher threshold
($E_{th}^{\psi'}-E_{th}^{J/\psi})\approx 3$ GeV the direct \psip production
off nuclei (impulse approximation) is suppressed compared to that of \jp by
the nuclear form factor at the same photon energy. Since $q_{\| }^{\gamma
\psi'}\approx {{m_{\psi'}^2}\over {m_{J/\psi}^2}} q_{\| }^{\gamma J/\psi} $
the minima of the impulse approximation distribution are shifted. The
contribution of the nondiagonal term $\gamma\to J/\psi\to \psi'$ essentially
increases the yield of the \psip and shifts the minima in the spectrum to
lower photon energies.  This results in significant effects, especially,
if one measure the relative $\psi' -to- J/\psi$ forward yield shown in 
fig.~\ref{ratfor}~\footnote{Because of the
significant amplitude for the \jp and \psip nondiagonal 
transitions the master matrix in eq.~(\ref{eqpsip})  governing the
$z$ dependence of the eikonal phase for the charmonium states in the nuclear
medium  is similar to that for the coherent 
 $K_L-K_S$ regeneration in the nuclear
medium. The similarity is enhanced by the fact that the \jp state dominates 
in 
the initial condition for these  equations because of the dominance of  
spatially 
small c$\bar c$ configurations  in the photoproduction processes and
significantly smaller size of \jp state. Thus the color screening phenomenon
reveal itself in the well familiar quantum mechanical phenomenon-oscillations 
between two states in the  medium.  The distinctive  feature of the charmonium
state production is the  large  (on nuclear scale) difference of  the
longitudinal momentum transfer  for the production of $J/\psi$ and $\psi'$
which leads to an oscillating energy dependence of the  $\psi'/J/\psi$
ratio. }.

To quantify the role of diagonal and nondiagonal rescatterings we presented
this ratio calculated in the Generalized Glauber model (solid line) and in
the impulse approximation (dotted line). Besides, to show separately the
influence of the diagonal and the nondiagonal transitions we show 
calculations in the GGM but with accounting for the nondiagonal transitions
only (dashed line, $f_{VN\to VN}=0$). The position of the minima
in the impulse approximation corresponds to the minima in the spectrum of
the \psip forward yield, position of maxima corresponds to minima in the \jp
yield. Easy to estimate that the energy shift is $\delta
\omega_{\gamma}\approx {R_A\over 7.6}(m_{\psi'}^2-m_{J/\psi}^2)$. The shape
is distorted by the energy dependence of the elementary photoproduction
amplitudes. The account of the nondiagonal transitions considerably shifts
the position of the minima and increases the relative yield. The diagonal
rescatterings produce some additional shift and essential modifications of
the shape:  at some energies one can find suppression, at others -enhancement.  
We emphasize that the measurement of such a ratio removes the nuclear model
dependence since the same longitudinal nuclear form factor enters the
numerator and the denominator. Hence, the ratio of cross sections in the
impulse approximation can be used as some kind of a model independent
reference curve since it can be easily calculated for many nuclei using the
nuclear form factors measured in a high energy elastic electron scattering
experiments and the ratio of the measured elementary photoproduction cross
sections. Besides, we would like to note that
one can also remove the dependence on the elementary photoproduction cross
sections measuring the double ratio of the relative $\psi'$-to-$J/\psi$
yield in the coherent photoproduction off two nuclei: a heavy nucleus 
and a light one.

In the case of a setup lacking a sufficient resolution in transverse
momentum one can study the energy dependence of the  $\psi'$ and
$J/\psi$ yields integrated over the transverse momentum. 
While such integration  smears the 
oscillation pattern significant oscillations still remain
as can seen in fig.~\ref{cs} where we compare the total cross sections
calculated in GGM to that in the Impulse Approximation. To emphasize the
significant influence of nondiagonal transitions we also show results
of calculation in the GGM without diagonal rescattering.

Finally, we want briefly comment the opportunity to extract the genuine 
$J/\psi N$ and $\psi' N$ cross section from such a measurement.
The conventional procedure is based on the estimate of the suppression of 
the particle yield comparing the data to the calculations within the Impulse 
Approximation and then describing them using a reasonable theoretical 
model for the process with the elementary cross sections used as 
fitting parameters. In the case of the
charmonium photoproduction such a procedure seems to be essentially 
complicated. While the sensitivity of the nuclear photoproduction
cross sections to the values of the elementary amplitudes is, in principle,
revealed in our calculations the interplay of diagonal and nondiagonal 
rescattering amplitudes
 and their interference with the amplitude of the direct production result
in rather formidable problem. In fig.~\ref{ratjpsi} we show the 
calculated ratio of the \jp photoproduction cross section in GGM to that 
in the IA. Despite the energy dependence of the \jp N cross section is
rather weak at the photon energies considered here  
we find the energy dependent suppression of the \jp yield
stronger $\approx (15\div 20)\%$ at $\omega \leq  40$ GeV and 
more or less small $\approx (6\div 7)\%$ at  $\omega \geq  40$ GeV in the
photoproduction off nucleus. Moreover, the increase of the elementary 
\jp N cross section by a factor 1.5 
that is allowed by the experimental uncertainties (see fig.~\ref{elcs})
changes the suppression by only $\approx 5\%$ at low energies and is
practically negligible at higher energies. Since $\sigma_{J/\psi N}$
is small one can easily
estimate that due to the \jp N interaction the suppression 
on the level of $\approx (30\div 40)\%$ should be expected
in the whole range of energy. Hence, even at low energies where  
the $\gamma N\to \psi' N$ amplitude is small
we find a noticeable compensation of
the suppression by contribution of the two-step 
$\gamma A\to A+\psi^{\prime}\to A+J/\psi$ production.
With increase of the photon energies in the considered region this 
compensation effect becomes stronger. This indicates 
moving to the regime of Color Transparency for a \jp passing through the
nuclear medium. The analysis of the
\psip photoproduction shows (fig.~\ref{ratpsi}) a significant influence of 
the two-step photoproduction 
$\gamma +A\to J/\psi +A\to \psi^{\prime} +A$ and the interference of 
this amplitude 
with the direct production and with the amplitude comprising diagonal 
rescattering in a wide range of energies.     
Hence, we can conclude that such a complicated interplay of rescatterings 
will preclude uniqueness of determining of the genuine \jp N and \psip N 
cross sections from measurements with one heavy nuclear target. Evidently
to succeed in this aim the elementary photoproduction and nondiagonal
amplitudes in a wide range of the photon energies should be known with
high precision. In our forthcoming papers we suggest a new approach to
this experimental problem and plan to analyze whether the study of the
A-dependence in the photoproduction of charmonium  
including the oscillation patterns
can be used to resolve the issue of determining the \jp N and \psip N
cross sections and the accuracy of the two diffractive state approximation
and hence provide further insights to the color dynamics at
the low energies. 

\section{Conclusion}
We have calculated the coherent charmonia photoproduction off heavy nuclei
at moderate energies within the Generalized Glauber Model which
combines the multistep production Glauber approach, the Generalized 
Vector Dominance Model and color screening phenomena.
We found that the interplay of the oscillating behavior of the nuclear form
factor and the interference of the rescattering amplitudes lead to
significant oscillations in the relative $\psi'$-to-$J/\psi$ yield
for forward scattering. We show
that accounting for the nondiagonal amplitudes, which model in the hadronic
basis the QCD color fluctuations within hadrons, results in the noticeable
modification of the coherent cross section of the \psip photoproduction
off nuclei. We found sensitivity of oscillations to the cross sections of
$J/\psi N$ and $\psi'$ N interaction as well as to the strength of the
nondiagonal transitions.
Accounting for the higher mass states may lead to larger cross section of 
$\psi'$N interactions $\approx 15$ mb, however, it will not 
change qualitatively the observed pattern of oscillations 
which is due to the oscillating behavior of the longitudinal nuclear 
form factor
at large (on the nuclear scale) longitudinal momentum transfer
and a large value of nondiagonal transitions expected 
as a basic feature of QCD.

For the coherent hidden beauty meson production at low energies the same 
oscillations in the $\Upsilon,\,\Upsilon'$ yields are expected but the  
cross sections will be obviously too small.

\vspace*{1cm}
{\bf Acknowledgment:}\\
This work was supported in part by GIF and DOE. L.G. thanks the Minerva 
Foundation for support.

\end{document}